\documentstyle[prl,multicol,aps,epsfig]{revtex}
\renewcommand{\widetext} 
{\end{multicols}\global\columnwidth42.5pc} 
\begin{document} 
\newcommand{\be}{\begin{equation}} 
\newcommand{\ee}{\end{equation}} 
\newcommand{\bea}{\begin{eqnarray}} 
\newcommand{\eea}{\end{eqnarray}} 
\newcommand{\br}{{\bf r}} 
\newcommand{\bk}{{\bf k}} 
\newcommand{\bq}{{\bf q}} 
\newcommand{\bn}{{\bf n}} 
\newcommand{\bp}{{\bf p}}

\draft 
\title{Interaction-induced magnetoresistance:
From the diffusive to the ballistic regime} 
\author{I.~V.~Gornyi$^{1,2,*}$ and A.~D.~Mirlin$^{2,1,\dagger}$} 
\address{$^1$ Institut f\"ur Theorie der Kondensierten Materie, 
Universit\"at Karlsruhe, 76128 Karlsruhe, Germany} 
\address{$^2$ Institut 
f\"ur Nanotechnologie, Forschungszentrum Karlsruhe, 76021 Karlsruhe, 
Germany} 
\date{\today} 
\maketitle 
\begin{abstract} 
We study 
interaction-induced quantum correction $\delta\sigma_{\alpha\beta}$ to 
the conductivity tensor of electrons in two dimensions 
for arbitrary $T\tau$, where 
$T$ is the temperature and $\tau$ the transport mean free time.  
A general formula is derived, expressing $\delta\sigma_{\alpha\beta}$ 
in terms of classical propagators (``ballistic diffusons'').  
The formalism is used to 
calculate the interaction contribution to the magnetoresistance in a 
classically strong transverse field and smooth disorder in the whole 
range of temperatures from the diffusive   
($T\tau\ll 1$) to the ballistic ($T\tau\gtrsim 1$) regime.  
\end{abstract} 
\pacs{PACS numbers: 
72.10.-d, 
73.23.Ad, 
71.10.-w, 
73.43.Qt  
} 
\begin{multicols}{2} 
\narrowtext 
 
\vspace{-0.3cm} 
The magnetoresistance (MR) in a transverse field $B$ is one of the 
most frequently studied characteristics of the two-dimensional (2D) 
electron gas 
 \cite{AA,Been}. 
Within the 
Drude-Boltzmann theory, 
the longitudinal resistivity of an isotropic degenerate system is 
$B$--independent, $\rho_{xx}(B)=\rho_0=(e^2\nu v_F^2 \tau)^{-1}$, 
where $\nu$ is the density of states per spin direction, 
$v_F$ the Fermi velocity, and $\tau$ the transport scattering time. 
There are several distinct sources of a non-trivial MR, 
which reflect the rich physics of 
2D systems. First, 
quasiclassical 
memory effects may lead to 
a MR~\cite{Antidots}, which shows no
$T$-dependence at low temperatures.
Second,
weak localization~\cite{AA} induces a negative MR
restricted to the range of very weak magnetic fields. 
Finally, another quantum correction to MR is generated by 
the electron--electron 
interaction. This effect 
is the subject of the present paper.

It was discovered by Altshuler and Aronov \cite{AA} 
that the Coulomb 
interaction enhanced by the diffusive motion of electrons 
gives rise to a quantum correction to conductivity, 
which has in 2D the form 
(we set $k_B=\hbar=1$) 
\be 
\delta\sigma_{xx} \simeq (e^2/2\pi^2) 
\ln T\tau, \; \qquad  T\tau \ll 1. 
\label{AAdiff} 
\ee 
It is assumed here for simplicity that $\kappa \ll k_F$,
where $\kappa=4\pi e^2\nu$ is 
the inverse screening length. 
The condition $T\tau\ll 1$ under which Eq.~(\ref{AAdiff}) 
is derived \cite{AA} implies that electrons move diffusively 
on the time scale $1/T$ 
and is termed the ``diffusive regime''. 
Subsequent works \cite{SenGir} showed that Eq.~(\ref{AAdiff}) 
remains valid in a strong magnetic field, leading (in combination with 
$\delta\sigma_{xy}=0$) to a parabolic 
interaction--induced quantum 
MR, 
\be 
{\delta\rho_{xx}(B)\over \rho_0} \simeq  
{{(\omega_c\tau)^2-1}\over \pi k_F l} \ln T\tau,  \; \qquad  T\tau \ll 1, 
\label{MRAA} 
\ee 
where $\omega_c=eB/mc$ is the cyclotron frequency and $l=v_F\tau$ the  
transport mean free path. Indeed, a $T$--dependent negative MR was 
observed in experiments \cite{PTH83Choi} and attributed to the  
interaction effect.  
However, the experiments~\cite{PTH83Choi} cannot be directly compared with
the theory \cite{AA,SenGir} since they were performed at higher
temperatures, $T\tau \gtrsim 1$.
(In high-mobility GaAs heterostructures conventionally
used in MR experiments, $1/\tau$ is typically $\sim 100\  {\rm mK}$
and becomes even smaller with improving quality of samples.) 
There is thus a clear need for a theory of the MR 
in the ballistic 
regime, $T \gtrsim 1/\tau$. 
 
In fact, the effect of interaction 
on the
conductivity 
at $T\gtrsim 1/\tau$ has 
attracted a 
great deal of interest  
in a 
context
of low-density 2D systems showing a seemingly  
metallic behavior, 
$d\rho/dT>0$~\cite{krapud}. 
Recently,
Zala, Narozhny, and Aleiner \cite{ZNA} developed 
a systematic theory of the interaction corrections 
valid for arbitrary $T\tau$. 
In the ballistic range of temperatures, this theory
(improving earlier calculation of temperature--dependent 
screening \cite{GD86}), 
predicts a linear-in-$T$ correction to  
conductivity $\sigma_{xx}$ and a $1/T$ correction to the Hall coefficient 
$\rho_{xy}/B$ at $B \to 0$, and describes the MR in a {\it parallel} field.

The consideration of \cite{ZNA} is restricted, however, 
to {\it classically weak} transverse fields,  
$\omega_c\tau \ll 1$, and to the {\it white-noise} disorder. 
The latter assumption is believed to be justified for Si-based 
and some p-GaAs structures, and the results of \cite{ZNA} have been 
by and large confirmed by most recent experiments~\cite{RecentMIT}
on such systems. 
On the other hand, the random potential in  
n-GaAs 
heterostructures 
is, as a rule, due to remote donors and has a long--range  
character. Thus, the impurity scattering is predominantly of a small--angle  
nature and is characterized by two relaxation times, the transport 
time $\tau$ 
and the single-particle (quantum) time $\tau_s$, 
with $\tau \gg \tau_s$.

We present here a general theory of the interaction--induced corrections
to the conductivity of 2D  
electrons valid for arbitrary temperatures, transverse 
magnetic fields and disorder range.   
We further apply it to the problem of 
magnetotransport 
in a smooth disorder at $\omega_c\tau \gg 1$~\cite{foot1}.
In the ballistic limit, $T\tau \gg 1$
(where the character of disorder is crucially important),
we show that while 
the correction to $\rho_{xx}$
is exponentially suppressed for $\omega_c\ll T$, 
a MR arises 
at stronger $B$ 
where it scales as $B^2T^{-1/2}$.

To find  $\delta\sigma_{\alpha\beta}$, we make use of the ``ballistic'' 
generalization of the Matsubara diffuson diagram technique of 
Ref.~\cite{AA}. We consider the exchange contribution first and will discuss  
the Hartree term later on. The relevant 
diagrams are shown in Fig.~1. 
The shaded blocks in Fig.~1 denote the impurity--line ladders, 
which we term ``ballistic diffusons''. The temperature range 
of main interest in the present paper is restricted by $T\tau_s \ll 1$, 
since at higher $T$ the MR will be small in the whole range of 
the quasiclassical transport  $\omega_c\tau_s \ll 1$ (see below). 
In this case the ladders are dominated by contributions with many 
($\gg 1$) impurity lines. Our general formula below is, however, valid  
irrespective of the value of $T\tau_s$. 
 \begin{figure} 
\narrowtext 
\centerline{ {\epsfxsize=6cm{\epsfbox{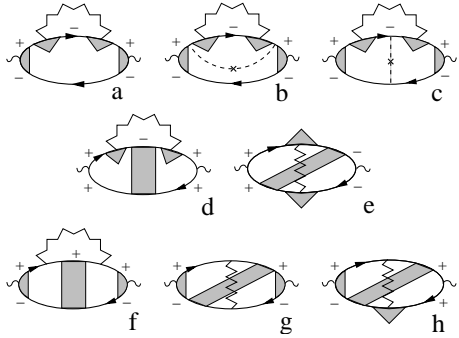}} }} 
\vspace{2mm} 
\caption{  Diagrams for the interaction  
correction to $\sigma_{\alpha\beta}$. 
The wavy (dashed) lines denote the interaction
(impurity scattering),
the shaded blocks are impurity ladders,
and the $+/-$ symbols denote the signs 
of the Matsubara frequencies. The diagrams obtained by a flip  
and/or by an exchange $+ \leftrightarrow -$ should also 
be included.} 
\label{fig1} 
\end{figure} 

After the Wigner transformation is performed, the ballistic diffuson 
takes the form ${\cal D}(\omega;\br,\bn;\br',\bn')$ and describes 
the quasiclassical propagation of an electron in the phase space 
\cite{RAG} ($\bn$ is the unit vector characterizing the direction of velocity 
on the Fermi surface). 
In contrast to the diffusive regime, where ${\cal D}$ has a universal and 
simple structure ${\cal D}(\omega,\bq)=1/(Dq^2-i\omega)$ 
determined by the diffusion constant $D$ only, its form in the  
ballistic regime is much more complicated. 
We are able, however, to get a general expression for  
$\delta\sigma_{\alpha\beta}$ in terms of the ballistic propagator 
${\cal D}(\omega,\bq;\bn,\bn')$. The results reads 
\begin{eqnarray} 
\delta\sigma_{\alpha\beta}&=&-2e^2 v_{\rm F}^2\nu \int_{-\infty}^\infty 
\frac{d\omega}{2\pi} 
\frac{\partial}{\partial \omega} 
\left\{\omega\  {\rm coth}{\omega\over{2T}}\right\}\nonumber \\ 
&\times&\int \frac{d^2{\bf q}}{(2\pi)^2}\  {\rm Im} 
\left[\ U(\omega,{\bf q})\  B_{\alpha\beta}(\omega,{\bf q})\ \right], 
\label{sigma} 
\end{eqnarray} 
where $U(\omega,{\bf q})$ is the interaction potential equal to 
a constant $V_0$ for point-like interaction and to 
\be 
U(\omega,{\bf q})= 
\frac{1}{2\nu}\ \frac{\kappa}{q+\kappa[1+i\omega\langle 
{\cal D}(\omega,q) \rangle]} 
\label{screen} 
\ee 
for screened Coulomb interaction. 
For
small--angle impurity scattering 
the tensor $B_{\alpha\beta}(\omega,{\bf q})$ in (\ref{sigma}) 
is given by 
\begin{eqnarray} 
B_{\alpha\beta}(\omega,{\bf q})&=& 
{T_{\alpha\beta} \over 2}\langle {\cal D}{\cal D}\rangle+ 
T_{\alpha\gamma}\left({\delta_{\gamma\delta}\over 2}\langle {\cal D}\rangle  
- \langle n_{\gamma} {\cal D} n_{\delta} \rangle \right) T_{\delta\beta} 
\nonumber \\ 
&-&2 T_{\alpha\gamma} \langle n_{\gamma} {\cal D} n_{\beta} {\cal D} \rangle  
- \langle {\cal D}  n_{\alpha} {\cal D} n_{\beta} {\cal D} \rangle, 
\label{Bwq} 
\end{eqnarray} 
where 
$
T_{\alpha\beta}=\left.\langle n_\alpha 
{\cal D}n_\beta\rangle\right|_{q=0,\omega\to 0}=
\sigma_{\alpha\beta}/e^2v_F^2\nu.
$
The angular brackets $\langle \dots \rangle$ in (\ref{screen}) and (\ref{Bwq})
denote averaging over velocity directions, e.g.               
$\langle n_x{\cal D}n_x\rangle=(2\pi)^{-2}\int d\phi_1 d\phi_2 
\cos\phi_1{\cal D}(\omega,\bq;\phi_1,\phi_2)\cos\phi_2,$ where $\phi$ is the 
polar angle of $\bn$. The first term in (\ref{Bwq}) originates 
from the diagrams {\it a,b,c} in Fig.~1 (forming together the Hikami box), 
the second term from {\it a,f,g}~\cite{footdrag}, 
the third term from {\it h}, 
and the last one -- from {\it d} and {\it e}. 
 
In the more general situation, when the 
scattering is at least partly of the large--angle character,  
the first term in (\ref{Bwq}) acquires a slightly more 
complicated form,  
\be 
\label{hikami} 
\pi\nu T_{\alpha\alpha'}[ 
\langle {\cal D} W{\cal D}\rangle S_{\alpha'\beta'} -  
2\langle {\cal D}n_{\alpha'} W n_{\beta'}{\cal D}\rangle]  
T_{\beta'\beta}, 
\ee  
where 
$W({\bf n},{\bf n'})$ is the  scattering cross-section 
and $S_{xx}=S_{yy}=1, \quad S_{xy}=-S_{yx}=\omega_c\tau_s$.
In particular, for the case of purely white-noise disorder (when 
$\tau=\tau_s$ and $W({\bf n},{\bf n'})=1/2\pi\nu\tau$) 
this yields   
${1 \over 2}T_{\alpha\beta}\langle {\cal D}\rangle \langle 
{\cal D}\rangle- \tau^{-1} T_{\alpha\alpha'}\langle {\cal D}n_{\alpha'} 
\rangle \langle n_{\beta'}{\cal D}\rangle T_{\beta'\beta}$.  
At $B=0$ we then recover (using the explicit form of the ballistic propagator 
for this case) the result for $\delta\sigma$ 
obtained in a different way in \cite{ZNA}. 
Needless to say, in the diffusive limit,  
we reproduce (for arbitrary $B$ and disorder range)  
the logarithmic correction (\ref{AAdiff}), (\ref{MRAA}) 
determined by the diagrams {\it a-e}. 
 
Before turning to the analysis of the results for the strong-$B$ 
regime, we consider briefly the $B=0$ case assuming the ballistic 
temperature range $T\tau\gg 1$. The structure of Eqs.~(\ref{sigma}), 
(\ref{Bwq}),  
(\ref{hikami}) implies that the interaction correction is governed by 
returns of a particle to the original 
point in a time $t\lesssim T^{-1}\ll \tau$. Such a quick return  
may be induced by a single back--scattering process, yielding the 
contribution $\delta\sigma_{xx}\sim e^2 \nu\tau W(2k_F)T\tau$. 
For the case 
of white-noise disorder this reduces to $\delta\sigma_{xx}\sim e^2 T\tau$, 
in agreement with \cite{GD86,ZNA}. However, in a smooth disorder with 
a correlation length $d\gg k_F^{-1}$ this contribution is suppressed 
by the factor $2\pi\nu\tau W(2k_F)\sim e^{-k_F d}$. The probability to 
return after many small-angle scattering events is also exponentially 
suppressed for $t\ll \tau$, yielding a contribution  
$\delta\sigma_{xx}\sim \exp[-{\rm const}(T\tau)^{1/2}]$. Thus, the 
interaction correction in the ballistic regime
is exponentially small at $B=0$ for the case of 
smooth disorder. Moreover, the same argument applies to the case of
a non-zero $B$, as long as $\omega_c \ll T$. 
 
The situation changes qualitatively in a strong magnetic field, 
$\omega_c\tau\gg 1$ and $\omega_c \gg T$. 
The particle experiences then within the time $t\sim T^{-1}$ multiple 
cyclotron returns to the region close to the starting point. The 
corresponding ballistic propagator satisfies the equation 
\bea 
\left[-i\omega + i v_{\rm F} q \cos\phi+\omega_c{\partial\over\partial\phi} 
\right. 
&-&\left.{1\over \tau}{\partial^2\over \partial \phi^2}\right] 
{\cal D}(\omega,q;\phi,\phi') 
\nonumber \\ 
&=& 
2\pi\delta(\phi-\phi'). 
\label{LB} 
\eea 
The approximate solution of (\ref{LB}) at $\omega_c\tau \gg 1$ has the form 
\bea 
&&{\cal D}(\omega,q;\phi,\phi')=\exp\{-iqR_c(\sin\phi-\sin\phi')\}
\nonumber \\
&&\times\left[ \frac{\chi(\phi)\chi(\phi')}{Dq^2-i\omega} 
+ \sum_{n \neq 0}  
\frac{e^{i n(\phi-\phi')}}{Dq^2-i(\omega-n\omega_c)+n^2/\tau} \right],
\label{diffuson} 
\eea 
where $\chi(\phi)=1-iqR_c\cos\phi/\omega_c\tau$ and
$D\simeq{R_c^2/2\tau}$ in strong $B$. Since characteristic 
frequencies in (\ref{sigma}) are $\omega\sim T\ll\omega_c$, 
it is sufficient to keep only the first term  
in square brackets in (\ref{diffuson}) to 
obtain the leading contribution. 
Then $\langle{\cal D}\rangle$ in 
(\ref{screen}) is given by 
\be 
\langle {\cal D} \rangle = J_0^2(qR_c)/(Dq^2-i\omega), 
\label{<D>} 
\ee 
where $J_0(x)$ is the Bessel function. 
Furthermore, combining all four terms in (\ref{Bwq}), we get 
\be 
B_{xx}(\omega,q)={J_0^2(qR_c)\over (\omega_c\tau)^2} 
\frac{D\tau q^2 }{(Dq^2-i\omega)^3}. 
\label{Bxx} 
\ee 
Note that Eqs.~(\ref{<D>}), (\ref{Bxx}) differ from those obtained in 
the diffusive regime by the factor ${J_0^2(qR_c)}$ only. 
This is related to the fact that the motion of the guiding center is 
diffusive even on the ballistic time scale $t\ll\tau$ (provided 
$t\gg\omega_c^{-1}$), while the additional factor 
corresponds to the averaging over the cyclotron orbit. 
 
\begin{figure} 
\narrowtext 
\centerline{ {\epsfxsize=3.6cm{\epsfbox{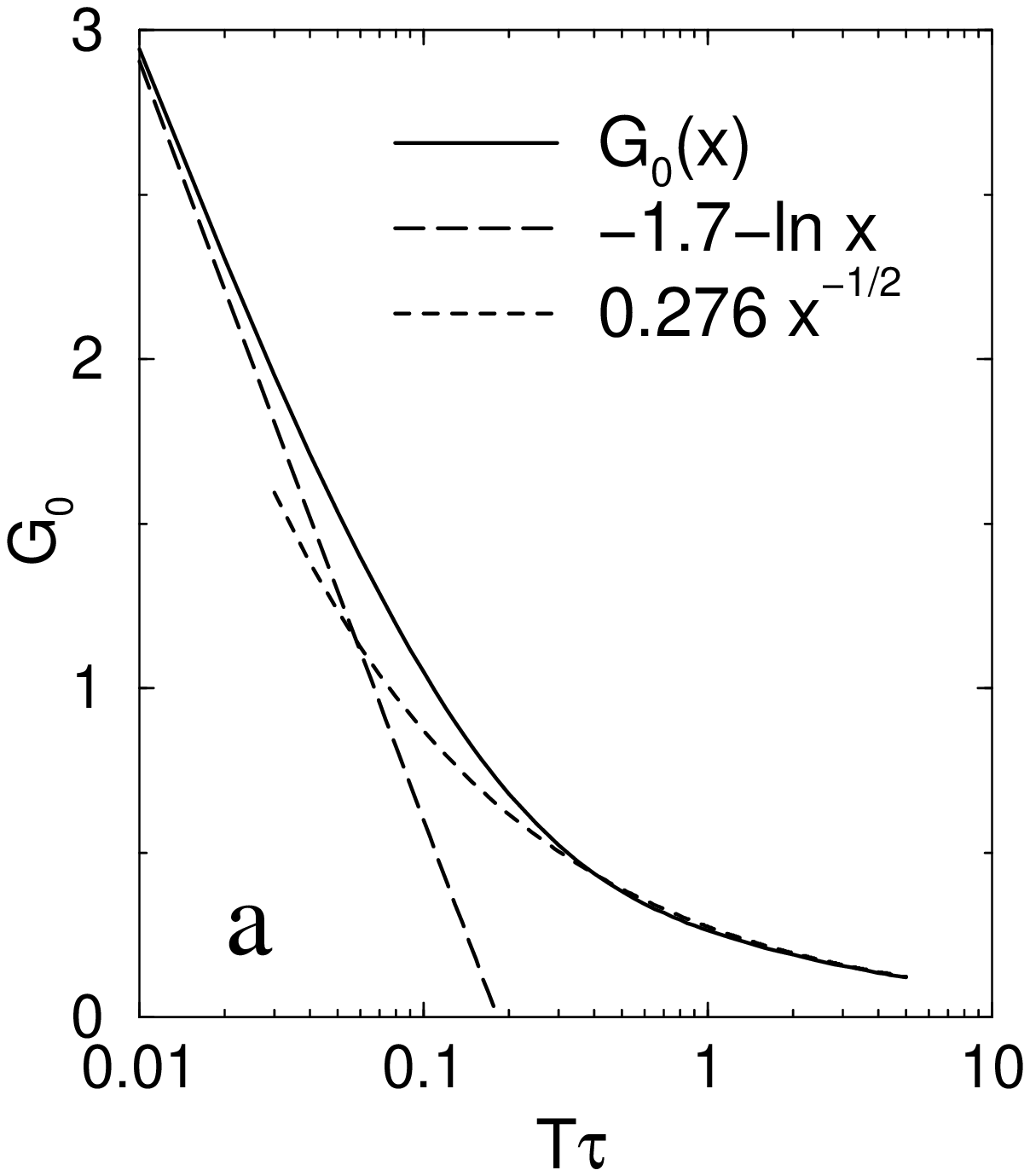}} }
{\epsfxsize=3.6cm{\epsfbox{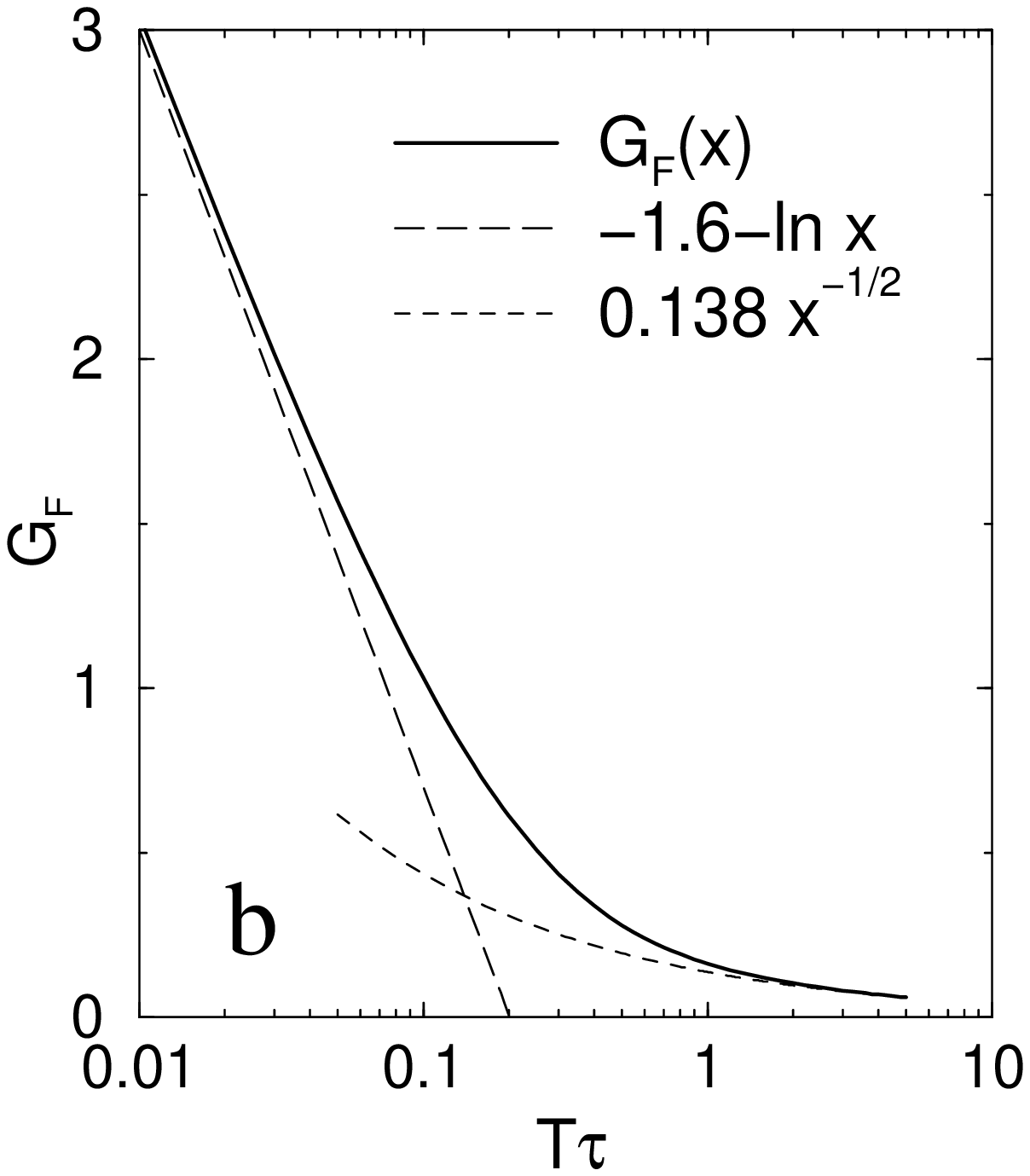}} }} 
\vspace{2mm} 
\caption{Functions $G_0(T\tau)$ (a) and $G_F(T\tau)$ (b)
determining the $T$-dependence of
the exchange 
term
for point-like, Eq.~(\ref{deltaint}), and Coulomb, Eq.~(\ref{exchange}),
interaction, respectively. 
} 
\label{fig2} 
\end{figure}

Substituting (\ref{Bxx}) in (\ref{sigma}), and rescaling
the momentum $q\to qR_c \equiv z$, we see that all
the $B$-dependence drops out from $\delta\sigma_{xx}$, 
and the exchange contribution in the case of point-like 
interaction reads 
\bea 
\delta\sigma_{xx}&=&-(e^2/2\pi^2)\nu V_0 G_0(T\tau),  
\label{deltaint} 
\\ 
G_0(x)&=&\pi^2 x^2 \!\! \int_0^\infty \!\!\!\!
\frac{d u \exp(-u)}{ u^3{\rm sinh}^2(\pi x/u)} 
\left[I_0(u)(1-u)+uI_1(u)\right].   
\nonumber 
\eea 
The Hartree term in this case is of the opposite sign and twice larger 
due to the spin summation (we neglect the Zeeman splitting). 
Since the relative correction to the Hall 
conductivity turns out to be smaller by the factor 
$\sim (\omega_c\tau)^{-2}$ compared to (\ref{deltaint}),  
$\delta\sigma_{xy}/\sigma_{xy} \ll \delta\sigma_{xx}/\sigma_{xx}$, the 
MR is given by 
$\delta\rho_{xx}/\rho_0=(\omega_c\tau)^2\delta\sigma_{xx}/\sigma_0$. 
The MR is thus quadratic in $\omega_c$, with the temperature 
dependence determined by the function $G_0(T\tau)$, 
which is shown in 
Fig.~\ref{fig2}a. It has the asymptotics $G_0(x)\simeq -\ln x+{\rm 
const}$ at $x\ll 1$ (diffusive regime) and $G_0(x)\simeq c_0x^{-1/2}$ 
with $c_0=3\zeta(3/2)/16\sqrt{\pi}\simeq 0.276$ at $x\gg 1$ (ballistic 
regime). Let us note that the crossover between the two limits takes 
place at numerically small values $T\tau\sim 0.1$.  

For the case of the Coulomb interaction the result turns out to be 
qualitatively similar. Substituting (\ref{screen}), (\ref{<D>}), and 
(\ref{Bxx}) in (\ref{sigma}), we get the exchange (Fock) 
contribution
\bea 
&&{\delta\rho^{\rm F}_{xx}(B)\over \rho_0} = 
-{(\omega_c\tau)^2\over \pi k_Fl} 
G_{\rm F}(T\tau), 
\label{exchange} \\
&&G_{\rm F}(x) =32 \pi^2 x^2
\int_0^\infty dz z^3 J_0^2(z) 
{\cal G}_{1,3,2}(z),
\nonumber \\ 
&&{\cal G}_{jkl}(z)=\sum_{n=1}^\infty  
\frac{n(12\pi xn[1-J_0^2(z)]+[3-j J_0^2(z)]z^2)} 
{(4\pi x n+z^2)^k (4 \pi x n[1-J_0^2(z)]+z^2)^l},
\nonumber 
\eea 
with $G_{\rm F}(x\ll 1)\simeq -\ln x 
+{\rm const}$ and $G_{\rm F}(x\gg 1)\simeq (c_0/2)x^{-1/2}$,
see Fig.~\ref{fig2}b.

\begin{figure} 
\narrowtext 
\centerline{ {\epsfxsize=4cm{\epsfbox{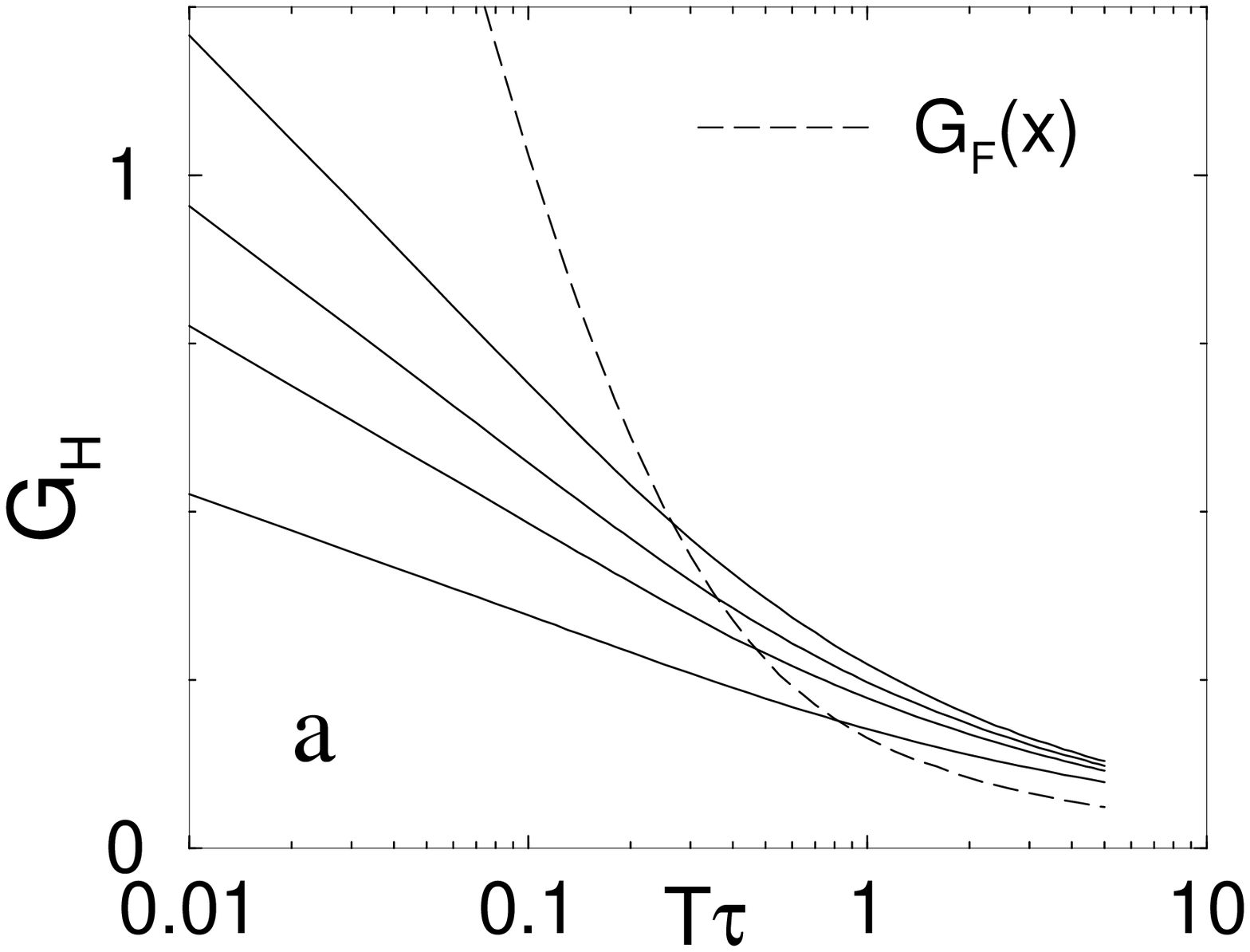}} } 
{\epsfxsize=4cm{\epsfbox{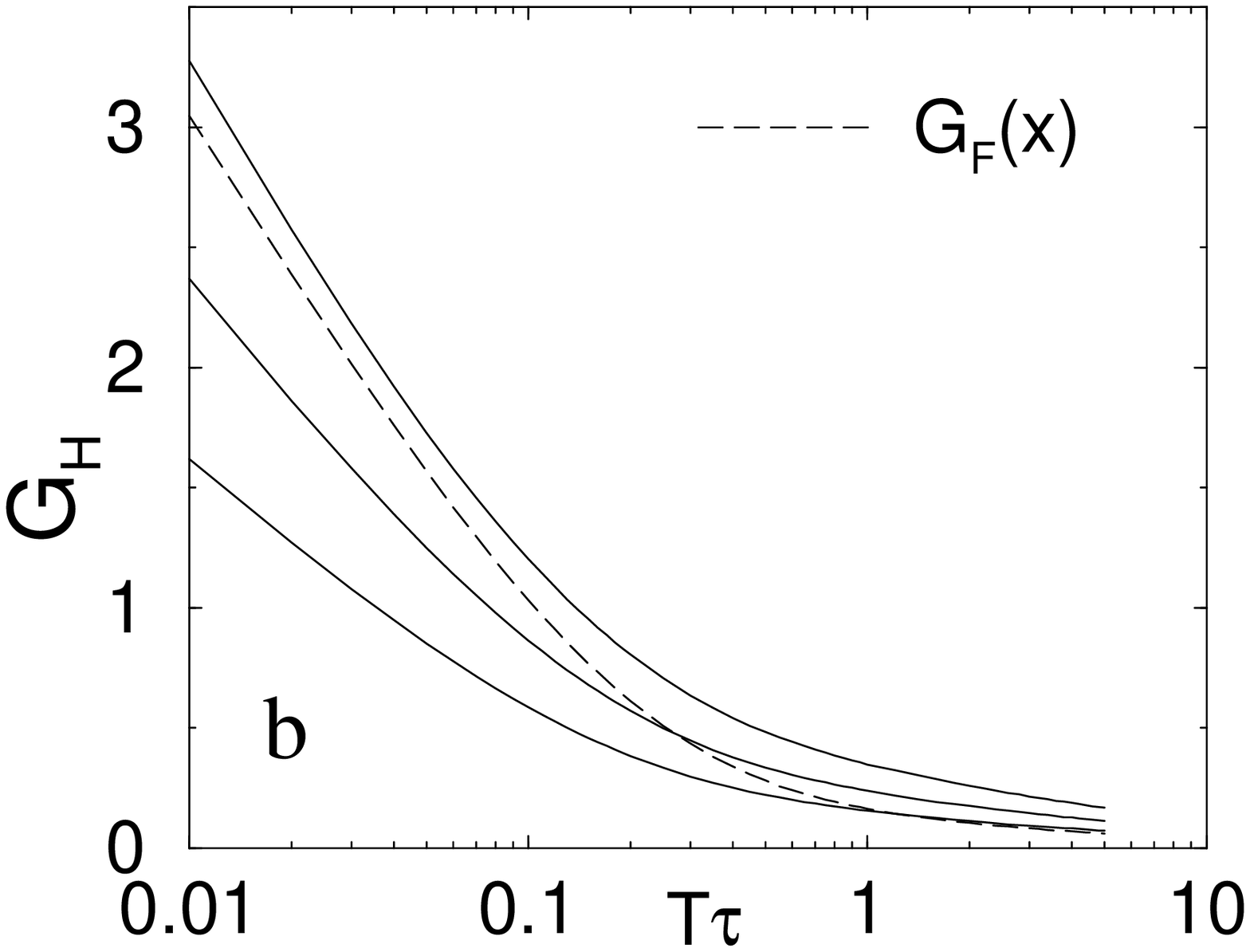}} }}
\vspace{0.5mm} 
\centerline{ {\epsfxsize=6.0cm{\epsfbox{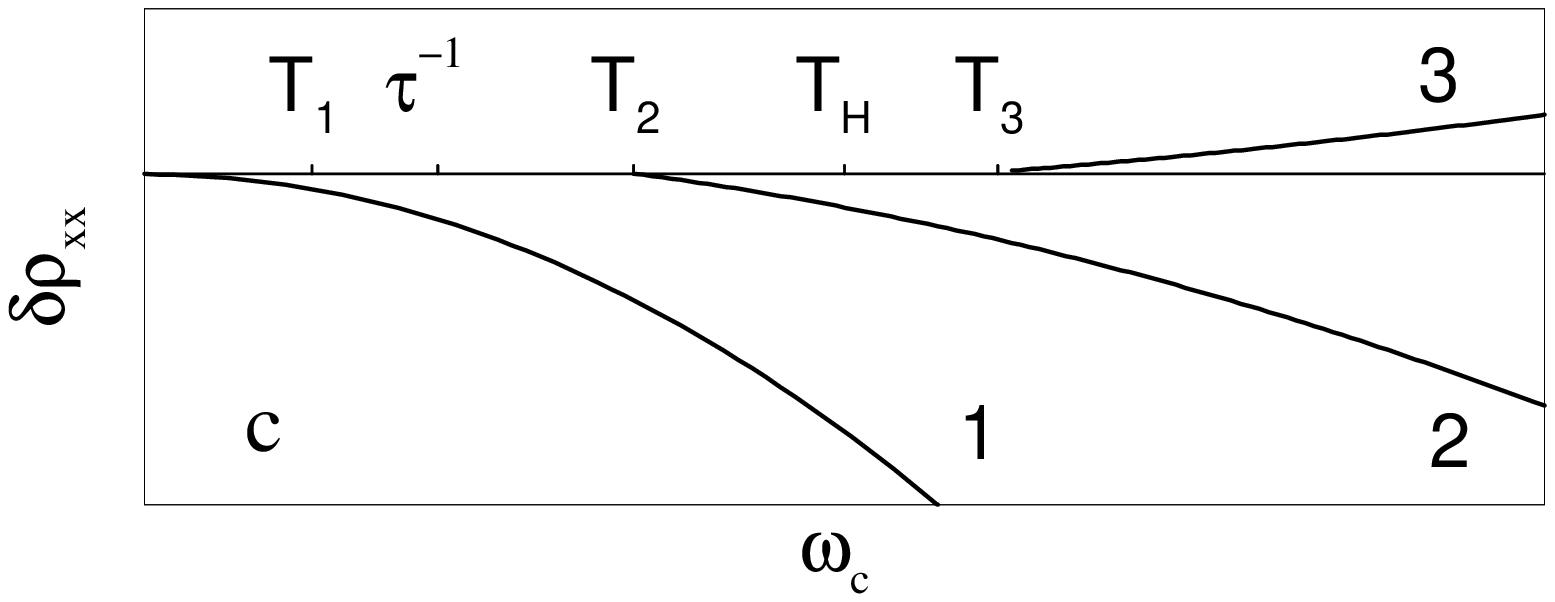}} }} 
\vspace{2mm} 
\caption{Hartree contribution, $ G_{\rm H}(T\tau)$, for
(a) weak interaction, $\kappa/k_F=0.1,\ 0.2,\ 0.3,\ 0.5$, and 
(b) strong interaction, $F_0=-0.3,\ -0.4,\ -0.5$
(from bottom to top);
(c) schematic plot of MR $\delta\rho_{xx}(B)$ 
in different temperature regimes:
1) $T_1\ll \tau^{-1}$, \ 2) $\tau^{-1}\ll T_2 \ll T_{\rm H}$, \ 
3) $T_3 \gg T_{\rm H}$. } 
\label{fig3} 
\end{figure}

We turn now to the Hartree 
term, assuming first 
$\kappa\ll k_F.$
The expression for its triplet part
is analogous to (\ref{sigma}) with the replacement of 
$U(\omega, \bq)$ by $-{3 \over 2}U(0,2k_F\sin[(\phi-\phi')/2])$, where 
$\phi$ and $\phi'$ are starting and final angles of the electron velocity.
As to the singlet 
part, 
it is renormalized by mixing with the exchange
term,
yielding
$$
U(\omega,q) \to
\frac{
\langle U(0,2k_F\sin{\phi-\phi' \over 2})\rangle 
-U(0,2k_F\sin{\phi-\phi' \over 2})}
{2[1+i\omega\langle {\cal D}(\omega,q) \rangle]^2}.
$$ 
(Note that the zero angular harmonic governing  
the diffusive limit~\cite{AA} is completely suppressed in 
the singlet part.)
After the 
angle integration,
$J_0^2(z)$ in (\ref{Bxx}) is replaced by 
$-(3y/2\pi)\int_0^\pi d\phi J_0(2z \sin\phi)/
(y+2\sin\phi)$ 
for the triplet, and
by ${\cal J}(y,z) = - (y/2\pi) \int_0^\pi d\phi 
[J_0(2z\sin\phi)-J_0^2(z)]/(y+2\sin\phi)$ 
for the singlet term ($y=\kappa/k_F$).
This yields for the total Hartree contribution
\bea
&&{\delta\rho^{\rm H}_{xx}(B)\over \rho_0}=
{(\omega_c\tau)^2\over \pi k_Fl}  
[G_{\rm H}^{\rm s}(T\tau,y)+
3G_{\rm H}^{\rm t}(T\tau,y)]
\label{rhohar} \\
&&\phantom{G_{\rm H}^{\rm t}}
\simeq {(\omega_c\tau)^2\over \pi^2 k_Fl}
\left\{\begin{array}{ll}  
y\ln y[{3\over 4}\ln(T\tau)+\ln y], & \ \ \ T\tau\ll 1,\\  
y\ln^2[y(T\tau)^{1/2}], &  1 \ll T\tau \ll y^{-2}, \\
{\pi c_0 (T\tau)^{-1/2}}, & \ \ \ T\tau \gg y^{-2},
\end{array}\right. 
\nonumber
\\
&&G_{\rm H}^{\rm s}(x,y)=32 \pi^2 x^2
\int_0^\infty dz z^3 {\cal J}(y,z) 
{\cal G}_{2,2,3}(z),
\nonumber \\
&&G_{\rm H}^{\rm t}(x,y)={\pi x^2\over 4 }\int_0^\infty \!\!\!
 \frac{du}{u^3{\rm sinh}^2(\pi x/u)}
 \int_0^\pi \! \! \! d\phi{y \over y+2\sin\phi} \nonumber \\
&&\phantom{G_{\rm H}^{\rm t}(x,y)}
\times \exp[-2u\sin^2\phi]
 (1-2u\sin^2\phi). 
 \nonumber
\eea 
We see that at $\kappa/k_F \ll 1$ a new energy scale 
$T_{\rm H} \sim \tau^{-1}(k_F/\kappa)^2$ arises where the MR changes sign.
Specifically, at $T \ll T_{\rm H}$ the MR,
$\delta\rho_{xx}=\delta\rho_{xx}^{\rm F}+\delta\rho_{xx}^{\rm H}$, 
is dominated
by the exchange term and is therefore negative,
while at
$T \gg T_{\rm H}$ the interaction becomes effectively point-like
and the Hartree term wins,
$\delta\rho_{xx}^{\rm H}=-2\delta\rho_{xx}^{\rm F}$,
leading to a positive MR
with the same $(T\tau)^{-1/2}$ temperature-dependence, 
see Fig. \ref{fig3}a,c.

If $\kappa/k_F$ is not small, 
the exchange contribution (\ref{exchange})
remains unchanged, while 
the Hartree term
is subject to strong Fermi-liquid renormalization 
\cite{AA,ZNA} and is determined by angular 
harmonics $F_m^{\sigma,\rho}$
of the Fermi-liquid interaction $F^{\sigma,\rho}(\theta).$ 
The formula for arbitrary $T\tau$ becomes then rather cumbersome 
\cite{unpub};
here we restrict ourselves to a discussion of limiting cases.
In the diffusive regime, 
$T\ll 1/\tau$, 
we reproduce
the known result~\cite{AA,ZNA}
$G_{\rm H}(T\tau)=3[1-\ln(1+F^\sigma_0)/F^\sigma_0]\ln T\tau$. 
In the ballistic limit,  $T\gg 1/\tau$,
we find for the Hartree contribution 
$$G_{\rm H}(T\tau)=-{c_0 \over 2}
\left[ 
\sum_{m\neq 0} {F^\rho_m \over 1+F^\rho_m}
+ 3 \sum_{m} {F^\sigma_m\over 1+F^\sigma_m} \right]
{1 \over \sqrt{T\tau}}.$$
Finally, within a frequently used (though parametrically uncontrolled)
approximation neglecting all $F_m$ with $m\neq 0$,
the Hartree term 
takes the form of Eq.~(\ref{exchange}) with
an additional overall factor of $3$ and with 
$J_0^2(z)$ multiplied by $F^\sigma_0/(1+F^\sigma_0)$ everywhere; 
the result is shown in Fig.~\ref{fig3}b for several values of 
$F^\sigma_0$.

In summary, we have derived a general formula for the 
interaction-induced quantum correction $\delta\sigma_{\alpha\beta}$ to 
the conductivity tensor of 2D  
electrons valid for arbitrary
temperature, magnetic field
and disorder range.  
It expresses $\delta\sigma_{\alpha\beta}$ in terms of 
classical propagators 
in random potential 
 (``ballistic diffusons''). Applying this 
formalism, we 
have calculated the interaction contribution to the MR 
in strong $B$ in a system with smooth disorder.  
We have shown that the parabolic MR found 
earlier in the diffusive 
linit
$T\tau\ll 1$ persists in the ballistic 
regime
$T\tau\gtrsim 1$, where it scales as $T^{-1/2}$. At 
sufficiently high $T$ the sign of the MR is changed (see Fig.~\ref{fig3}c).  
 
Before closing the paper, we list a few further applications of our 
formalism
\cite{unpub}. First, we can consider the model of mixed disorder, in 
which $\tau_s$ is determined by a smooth random potential while $\tau$ 
is governed by rare short-range scatterers. This model is relevant to 
ultra-high mobility heterostructures as well as to random antidot 
arrays~\cite{Antidots}. Second, the 
interaction correction to the MR in a periodically modulated system 
(lateral superlattice) can be studied.  
Finally, our results 
can be generalized to frequency-dependent MR.   

After completion of this work, we learnt about a 
recent experiment~\cite{Sav} supporting our theoretical 
predictions. We thank A.K.~Savchenko for useful discussions
and for informing us about results of~\cite{Sav} prior to
publication. 
This work was supported by the Schwerpunktprogramm 
``Quanten-Hall-Systeme'' and the SFB195 der Deutschen 
Forschungsgemeinschaft, and 
by the RFBR.
 
\vspace{-0.3cm}  


\end{multicols}


\begin{references}
\bibitem{foot1} 
A related problem of the tunneling density of states
in this situation was studied in Ref.~\cite{RAG}.
\end{references}
\end{document}